\documentclass[10pt,conference]{IEEEtran} 
\IEEEoverridecommandlockouts

\usepackage{cite}
\usepackage{amsmath,amssymb,amsfonts}
\usepackage{graphicx}
\usepackage{textcomp}
\usepackage{xcolor}
\def\BibTeX{{\rm B\kern-.05em{\sc i\kern-.025em b}\kern-.08em
    T\kern-.1667em\lower.7ex\hbox{E}\kern-.125emX}}
\usepackage{booktabs}
\usepackage{xcolor}
\usepackage[most]{tcolorbox}
\usepackage{tabularx}
\usepackage{soul}
\usepackage{hyperref}
\usepackage{multirow}
\usepackage{fancybox}
\usepackage{enumitem}
\usepackage{balance}
\usepackage{booktabs}
\usepackage{color}
\usepackage{float}
\usepackage{xcolor}
\usepackage{listings}
\usepackage{tabularx}
\usepackage{adjustbox}
\usepackage{array}
\usepackage{amsmath}
\usepackage{xspace}
\usepackage{url}
\usepackage{tikz}
\usepackage{caption}
\usepackage{color}
\usepackage{graphicx}
\definecolor{dkgreen}{rgb}{0,0.6,0}
\definecolor{gray}{rgb}{0.5,0.5,0.5}
\definecolor{mauve}{rgb}{0.58,0,0.82}
\usepackage{pgf-pie}
\usepackage{algorithm}
\usepackage{algpseudocode}
\usepackage{subcaption}

\usepackage{tcolorbox}
\lstdefinestyle{CppStyle}{
  language=C++,
  basicstyle=\ttfamily\footnotesize,
  keywordstyle=\color{blue},
  commentstyle=\color{green!50!black},
  stringstyle=\color{red},
  numbers=left,
  numberstyle=\tiny\color{gray},
  stepnumber=1,
  breaklines=true,
  tabsize=2,
  showspaces=false,
  showstringspaces=false
}

\lstdefinestyle{PythonStyle}{
  language=Python,
  basicstyle=\ttfamily\footnotesize,
  keywordstyle=\color{blue}\bfseries,
  commentstyle=\color{green!50!black},
  stringstyle=\color{red},
  numbers=left,
  numberstyle=\tiny\color{gray},
  stepnumber=1,
  breaklines=true,
  tabsize=4,
  showspaces=false,
  showstringspaces=false
}

\makeatletter
\newcommand{\mybox}[1]{%
	\setbox0=\hbox{#1}%
	\setlength{\@tempdima}{\dimexpr\wd0+13pt}%
	\begin{tcolorbox}[boxrule=0.5pt, colback=white, arc=4pt,
		left=6pt,right=6pt,top=6pt,bottom=6pt,boxsep=0pt]
		#1
	\end{tcolorbox}
}
\usepackage[framemethod=default]{mdframed}
\definecolor{lightblue}{RGB}{227,242,253}
\definecolor{codegreen}{rgb}{0,0.6,0}
\definecolor{codegray}{rgb}{0.5,0.5,0.5}
\definecolor{codepurple}{rgb}{0.58,0,0.82}
\definecolor{backcolour}{rgb}{0.95,0.95,0.92}

\lstdefinestyle{mystyle}{
  language=Python,
  aboveskip=3mm,
  showstringspaces=false,
  columns=flexible,
  numbers=none,
  backgroundcolor=\color{backcolour},
  commentstyle=\color{codegreen},
 keywordstyle=\color{magenta},
    numberstyle=\tiny\color{codegray},
    stringstyle=\color{codepurple},
    basicstyle=\small\ttfamily,
    breakatwhitespace=false,         
    breaklines=false,                 
    captionpos=b,                    
    keepspaces=false,                 
    numbersep=5pt,                  
    showspaces=false,                
    showstringspaces=false,
    showtabs=false,                  
    tabsize=2,
    escapeinside=``
}
\definecolor{nima}{RGB}{1.0, 0.49, 0.0}
\definecolor{songcolor}{RGB}{191,191,191}
\definecolor{nimacolor}{RGB}{0.13, 0.67, 0.8}
\definecolor{aruncolor}{RGB}{51,255,51}

\newmdenv[
  backgroundcolor=lightblue,
  linecolor=black,
  skipabove=5pt,
  skipbelow=5pt,
  linewidth=1pt,
  innertopmargin=3pt,
  innerbottommargin=3pt,
  innerleftmargin=3pt,
  innerrightmargin=3pt,
  frametitlebackgroundcolor=lightblue,
  frametitlefont=\bfseries,
]{findingbox}

\newmdenv[
  backgroundcolor=white,
  linecolor=black,
  skipabove=8pt,
  skipbelow=8pt,
  linewidth=0.6pt,
  roundcorner=3pt,
  innertopmargin=5pt,
  innerbottommargin=5pt,
  innerleftmargin=6pt,
  innerrightmargin=6pt,
]{exampleframe}

\definecolor{rescuegreen}{RGB}{46,125,50}
\definecolor{rescuebg}{RGB}{232,245,233}
\definecolor{regressred}{RGB}{198,40,40}
\definecolor{regressbg}{RGB}{253,235,234}
\newcommand{\rrcell}[4]{% #1 rule color  #2 bg color  #3 title  #4 body
  {\setlength{\fboxsep}{5pt}%
  \noindent\fcolorbox{#1}{#2}{%
    \begin{minipage}{\dimexpr\linewidth-2\fboxsep-2\fboxrule\relax}%
      \raggedright
      {\bfseries\scriptsize\textcolor{#1}{#3}\par}\smallskip%
      \scriptsize #4\par
    \end{minipage}%
  }}%
}
\newcommand{\rescueregress}[6]{%
  \begin{figure}[!htb]
    \centering
    \begin{minipage}{\linewidth}%
      \rrcell{rescuegreen}{rescuebg}{#3}{#4}%
      \par\smallskip
      \rrcell{regressred}{regressbg}{#5}{#6}%
      \par\smallskip
      \refstepcounter{figure}\label{#1}%
      \vspace{0.1cm}
      {\normalsize Figure~\thefigure: #2\par}%
    \end{minipage}%
  \end{figure}%
}

\begin{document}

\title{How Should Agents See Code? Rethinking Code Representations for Agentic Software Engineering}

%\author{Anonymous Authors}
\author{%
\IEEEauthorblockN{Melika Sepidband}
\IEEEauthorblockA{\textit{York University}\\
Toronto, Canada\\
melikasp@yorku.ca}
\and
\IEEEauthorblockN{Jiho Shin}
\IEEEauthorblockA{\textit{York University}\\
Toronto, Canada\\
shinjiho@yorku.ca}
\and
\IEEEauthorblockN{Song Wang}
\IEEEauthorblockA{\textit{York University}\\
Toronto, Canada\\
wangsong@yorku.ca}
\and
\IEEEauthorblockN{Hadi Hemmati}
\IEEEauthorblockA{\textit{York University}\\
Toronto, Canada\\
hemmati@yorku.ca}
}

\maketitle

\begin{abstract}
LLM agents are increasingly used for repository-level software engineering tasks. These tasks require agents to retrieve the right code context before reasoning or editing. However, most agents still access repositories through raw files, lexical search, and shell commands, treating code primarily as text. This paper investigates map-based code-representation workflows as a design variable in agentic software engineering, examining how different repository views change agents' retrieval behavior, task success, token use, and latency.

In this paper, we study five code-representation workflows: abstract syntax trees, call graphs, module dependencies, documentation-enriched search, and a hybrid workflow that combines them. Each workflow exposes an on-demand repository map that the agent can query during a task. We evaluate Claude Code with Haiku~4.5 against a raw-tool baseline on four repository-level tasks: fault localization with LinuxFLBench, repository-level code generation and repair with SWE-bench Pro, test generation with SWT-bench, and security patching with SecBench. We measure task success together with representation-tool use, native tool use, token consumption, and wall-clock time.

The results show that representation workflows are most useful as token-saving tools when localization is the bottleneck, while their success gains narrow when the decisive step is patch or test authoring. On LinuxFLBench, all five workflows improve recall@10 over the baseline (0.757--0.829 vs. 0.717) and reduce token usage by 25--39\%, but they also increase wall time. On SWE-bench Pro, resolution remains close to the baseline (44.3--50.6\% vs. 49.0\%) while token usage drops by 17--26\%. On SecBench, Callgraph has the highest patch-resolution rate in this sample (34.3\% vs. 28.9\%) and uses fewer tokens, but all workflows increase latency. On SWT-bench, Hybrid and Docs have the highest reproducing-test rates in this sample (47.8\% and 47.2\% vs. 44.7\%), but all workflows use more tokens and take longer. Overall, map-based workflows change both what the agent inspects and how much it spends, but their benefits are task-dependent rather than universal.
\end{abstract}


\section{Introduction}
\label{sec:intro}

LLM agents are increasingly used for repository-level software engineering (SE) tasks, such as fault localization, issue resolution, test generation, and security patching \cite{fan2023large,hou2024large,wang2025agents}. In these tasks, generating code is only one part of the process. The agent first needs to identify where the relevant behavior is implemented, which files and functions are related, and what evidence should be inspected before making a prediction or edit. If this localization step is weak, the agent may spend most of its budget reading irrelevant files or editing code that is only lexically related to the task.

Current coding agents, including Claude Code~\cite{anthropic2026claudecode}, SWE-agent~\cite{yang2024sweagent}, Agentless~\cite{xia2025demystifying} and OpenHands~\cite{wang2025openhands}, typically interact with repositories through lexical file search and shell-based exploration commands. This interface is flexible, but it gives the agent little structured information about the repository. For example, the agents may need to repeatedly search and read files to determine where a symbol is defined, which functions call it, which modules depend on it, or which documentation explains its intended behavior. These operations primarily serve to reconstruct the context required for the SE task. 

Code representations naturally expose this context. Beyond source text, a repository can be viewed through abstract syntax trees, call graphs, dependency graphs, documentation linked to program symbols, or combinations of these views. Each view emphasizes different relations: abstract syntax trees expose definitions and containment; call graphs expose inter-procedural relationships; dependency graphs expose module-level coupling; and documentation connects natural-language descriptions to code intent. These views are commonly used in program analysis and code intelligence \cite{mou2016convolutional, allamanis2018learning, guo2021graphcodebert, zhou2023docprompting, alon2019code2vec}, but their role in LLM agent workflows remains poorly understood.

Prior work on SE agents has shown that repository navigation and context selection affect performance \cite{yang2024sweagent, zhang2024autocoderover, yu2025orcaloca,wu2024repoformer, liu2024graphcoder, ouyang2025repograph}. However, representation is often bundled with the agent scaffold, retrieval method, or repair loop. This makes it difficult to determine whether an improvement comes from the model, the prompt, the agent policy, or the code view provided to the agent. As a result, we lack a systematic comparison of which representations help agents, for which tasks, and at what cost.

%\jiho{In this paper, we set out to investigate how agents should access code representations during repository-level software engineering tasks. We focus on SE tasks including fault localization, program repair, security patching, and test generation to cover a diverse set of repository-level workflows, outputs, and success criteria. These tasks stress different context bottlenecks: locating relevant code without editing, editing and validating a functional fix, repairing vulnerability-triggering behavior, and writing tests that expose a bug.}
%\song{here say what tasks you explore, and the reasons why you choose these tasks}

% We expose each representation as a Model Context Protocol (MCP) server rather than as a static prompt context. Each server builds a repository map once and provides query tools that return ranked, source-backed code neighborhoods. This design keeps the agent and model the same, while changing only the map shown to the agent and recording the map-use guidance used in the runs.\song{this part is implemntation details not necessary to show here} \jiho{removed}

\begin{figure}[t!]
    \centering    
    \includegraphics[width=\linewidth]{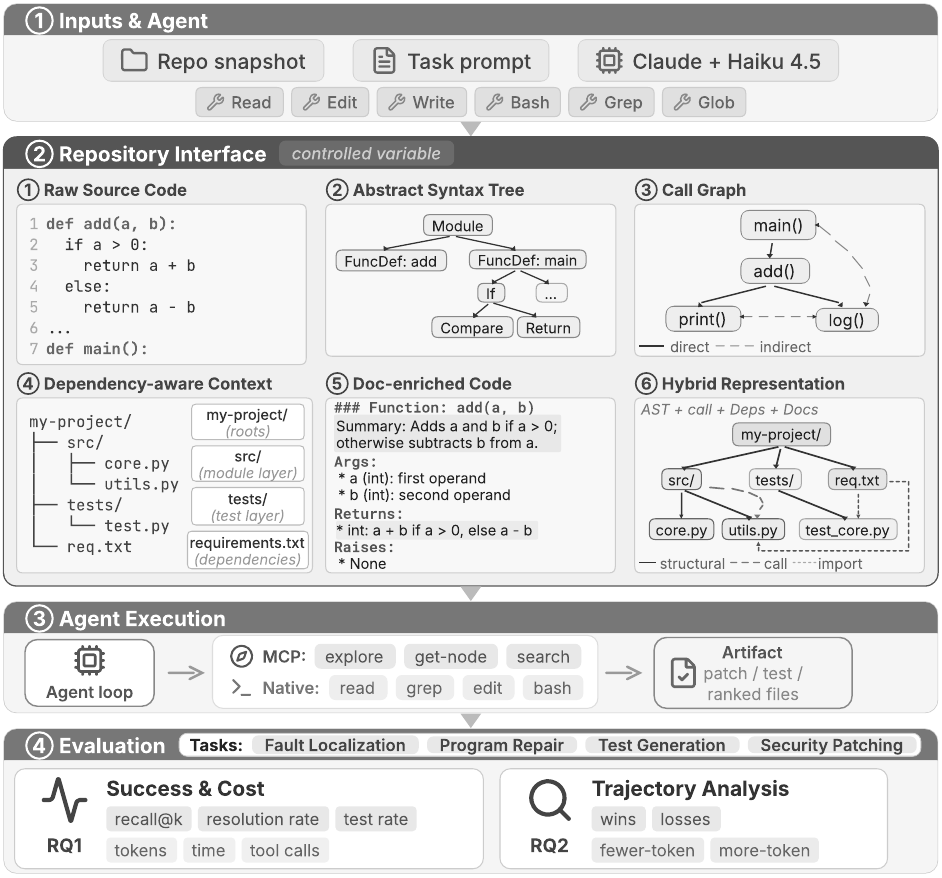}
    \caption{Overview of the proposed study.}
    \label{fig:1}
\end{figure}

To bridge the gap, in this paper, we investigate how agents should access code representations during repository-level software engineering tasks.
Figure~\ref{fig:1} shows an overview of our study. We study four repository-level software engineering tasks that differ in the repository understanding they require: fault localization (finding the files or symbols responsible for a bug from its report), program repair (editing code across a repository to fix a reported problem), security patching (locating and repairing the code responsible for a vulnerability), and test generation (understanding the behavior to exercise and writing a test that captures it). We instantiate these tasks with established repository-level benchmarks: LinuxFLBench~\cite{zhou2026taming}, SWE-bench Pro~\cite{deng2025swe}, SecBench~\cite{lee2026sec}, and SWT-bench~\cite{mundler2024swt}. These tasks stress different bottlenecks: locating relevant code without editing, editing and validating a functional fix, repairing vulnerability-triggering behavior, and writing tests that expose a bug.
We examine four representation families: abstract syntax trees (AST), call graphs (CG), dependency graphs (Deps), and documentation-enriched search (Docs). We also study a hybrid workflow that that combines these four views in a single repository map. For our evaluation, we measure task success, tool-call behavior, token usage, wall-clock time, and per-instance trajectories.

Our central finding is that map-based representation workflows are most useful when the task bottleneck is finding relevant code. On fault localization, where the task is code search, they improve recall while cutting token usage, although they add latency. On program repair, the bottleneck is partly past localization, so they preserve resolution while reducing tokens by replacing repository browsing. On security patching, the strongest signal in this sample comes from Callgraph, which matches stack-centered prompts and exposes caller/callee context at lower token cost but higher latency. On test generation, where the decisive step is authoring a discriminating assertion, success changes are small, and several workflows add token overhead. Thus, the results support a task-dependent workflow claim rather than a universal claim that any fixed representation is better.
%Overall, code representations are most valuable when they replace repository localization work, and their payoff depends on the task bottleneck.

\smallskip
\noindent
This work makes the following contributions:
\begin{itemize}[itemsep=2pt,topsep=2pt,leftmargin=*]
    \item To our knowledge, we provide the first controlled comparison of five code data representations for agentic software engineering, evaluating AST, call graph, dependency, documentation-enriched, and hybrid views under the same agent stack across four repository-level tasks.
    \item We compare AST, call graph, dependency, documentation-enriched, and hybrid repository views under a fixed agent stack, reporting task success together with token usage, wall-clock time, and tool-call behavior.
    \item We show that representation workflows provide their strongest and most consistent benefit when repository localization is the primary bottleneck, mainly by reducing token-heavy browsing.
    \item We provide trajectory-level evidence explaining when different repository views help, add overhead, or fail to affect the final patch or test-generation step.
\end{itemize}

\noindent \textbf{Data Availability:} 
We provide the representation servers, index and map builders, agent conditions, experiment harness, and metrics scripts online to facilitate replication and future work at: %\url{https://anonymous.4open.science/r/How-Should-Agents-See-Code-5883}
\url{https://github.com/MelikaSepidband/How-Should-Agents-See-Code}

\section{Related Work}
\label{sec:related_work}

\subsection{Repository-Level Agentic Software Engineering}
Repository-level SE agents often search, edit, execute, and validate code across an entire project. SWE-bench \cite{jimenez2024swe} established this setting with real GitHub issues that require cross-file reasoning, while SWE-agent \cite{yang2024sweagent} showed that the repository interface shapes how agents read files, edit code, and run commands. AutoCodeRover \cite{zhang2024autocoderover} emphasizes localization before repair through structure-aware code-entity search. Together, these studies show that repository navigation is part of the task, not merely a pre-processing step.

Recent systems improve agentic software engineering by modifying the agent workflow. Agentless \cite{xia2025demystifying} decomposes issue resolution into localization, repair, and validation, showing that simple scaffolds can be effective when context is carefully selected. OrcaLoca \cite{yu2025orcaloca} improves issue localization through action scheduling, decomposition, and context pruning, while ARISE \cite{seddik2026arise} exposes repository graphs and slicing tools for localization and repair. These works motivate our study, but they mainly evaluate complete systems or specific representation designs; in contrast, we keep the agent harness fixed and compare representations as the experimental variable.

\subsection{Code Representations for Structure and Intent}
Code intelligence research has long represented programs with views beyond local tokens. TBCNN \cite{mou2016convolutional} modeled syntactic structure with ASTs, Allamanis et al. \cite{allamanis2018learning} represented programs as graphs with semantic relations, and GraphCodeBERT \cite{guo2021graphcodebert} incorporated data-flow into pre-training. Documentation provides a complementary natural-language view of intent and API usage: DocPrompting \cite{zhou2023docprompting} retrieves documentation for code generation, and RepoRepair \cite{pan2026reporepair} uses hierarchical documentation for repository-level repair. Together, these studies show that syntax, program relations, and documentation each expose context that raw source tokens may not make explicit.

We study these representations in a different setting. Rather than embedding them inside a model, retriever, or repair system, we expose AST, call-graph, dependency, documentation, and hybrid views as MCP tools during an agent trajectory. This lets us compare which view helps which repository-level task under a shared interface and fixed agent harness.

\subsection{Repository-Level Context Retrieval}
Repository-level coding requires selecting context before generation. RepoBench \cite{liu2024repobench} and CrossCodeEval \cite{ding2023crosscodeeval} evaluate completion when context spans files, while RepoCoder \cite{zhang2023repocoder} uses iterative retrieval and generation. Repoformer \cite{wu2024repoformer} shows that more context is not always better, and RepoQA \cite{liu2024repoqa} shows long-context models still need to connect natural-language queries with distributed code evidence.

Graph-based retrieval offers another way to organize repository context. GraphCoder \cite{liu2024graphcoder}, RepoGraph \cite{ouyang2025repograph}, LocAgent \cite{chen2025locagent}, and CodeGraph \cite{codegraph} expose repository graphs or graph-guided search tools for code navigation and localization. These systems show the value of interactive graph-based context, but each commits to one graph design and retrieval policy; our study instead holds the agent fixed and treats the code view as the controlled variable, comparing AST, call graph, dependency, documentation, and hybrid maps under one interface.
\section{Methodology}

\subsection{Code Representations}
\label{sec:approach:representations}

\begin{table}[t]
\centering
% \caption{The five code-representation servers and the MCP tools they expose to the agent. Every server exposes a shared \texttt{explore}/\texttt{node} navigation pair (and generic \texttt{files}/\texttt{status} utilities) over its own precomputed map, plus the representation's specialized tools.}
\caption{MCP tools exposed for each code-representation.}
\label{tab:reps}
\resizebox{\columnwidth}{!}{%
\begin{tabular}{l c >{\raggedright\arraybackslash}p{5.2cm}}
\toprule
Server & \#tools & Tools exposed \\
\midrule
\textbf{AST} (structural) & 8 &
\texttt{ast\_explore}, \texttt{ast\_node}, \texttt{ast\_search},
\texttt{ast\_files}, \texttt{ast\_status}, \texttt{file\_skeleton},
\texttt{ts\_query}, \texttt{parse\_file} \\
\textbf{Callgraph (CG)} (inter-procedural flow) & 9 &
\texttt{callgraph\_explore}, \texttt{callgraph\_node},
\texttt{callgraph\_search}, \texttt{callgraph\_files},
\texttt{callgraph\_status}, \texttt{callers\_of}, \texttt{callees\_of},
\texttt{call\_path}, \texttt{function\_info} \\
\textbf{Deps} (module dependency graph) & 8 &
\texttt{deps\_explore}, \texttt{deps\_node}, \texttt{deps\_files},
\texttt{deps\_status}, \texttt{module\_deps}, \texttt{dependency\_path},
\texttt{layer\_summary}, \texttt{cycles} \\
\textbf{Docs} (documentation-enriched) & 6 &
\texttt{docs\_explore}, \texttt{docs\_node}, \texttt{docs\_files},
\texttt{docs\_status}, \texttt{module\_overview}, \texttt{repo\_map} \\
\textbf{Hybrid} (composite) & 6 &
\texttt{hybrid\_explore}, \texttt{hybrid\_node}, \texttt{hybrid\_files},
\texttt{hybrid\_status}, \texttt{impact\_of}, \texttt{repo\_map} \\
\bottomrule
\end{tabular}}
\end{table}

We examine five different code representations, each capturing a distinct view of repository structure and intent. The details of each representation are described below.

\paragraph{AST} The abstract syntax tree captures the \emph{structural organization} of the code: which definitions (functions, methods, classes, structs, variables) exist and which are nested inside which, independent of runtime behavior. The AST server's map has these definitions as nodes and \emph{syntactic containment} as edges (file\,$\to$\,top-level definition\,$\to$\,nested definition), so the repository becomes a forest of containment trees rooted at files. This tells the agent where the code lives and how it is grouped. Its \texttt{explore}/\texttt{node} pair ranks definitions by structural relevance and returns a definition's source with its enclosing scope and nested members; additional tools (Table~\ref{tab:reps}) expose file skeletons and precise tree-sitter queries. In effect, AST makes the code's structural skeleton directly addressable without the agent re-parsing the source.

\paragraph{Callgraph (CG)} The call graph captures \emph{behavioral} structure: not where code is defined, but how control flows between definitions, so functions in different files are adjacent when one calls the other. The CG map has functions and methods as nodes and resolved calls (caller\,$\to$\,callee) as edges, forming a directed who-calls-whom graph. Its \texttt{explore}/\texttt{node} pair ranks functions by call-connectivity and returns a function's source with its callers and callees; further tools (Table~\ref{tab:reps}) return immediate callers/callees and concrete call paths. This view suits tasks where a symptom in one function must be traced to its cause in another along call edges.

\paragraph{Deps} The dependency graph captures structure at a coarser grain: not how definitions nest or call one another, but how whole files and modules rely on each other, so two files are adjacent when one imports the other. The deps server's map has files as nodes (with their definitions attached, so exploration surfaces real code rather than bare file names) and imports (file\,$\to$\,imported file) as edges. Its \texttt{explore}/\texttt{node} pair ranks code by import-distance to the query and returns a file's relevant code with the files it imports and is imported by; further tools (Table~\ref{tab:reps}) report a file's inbound/outbound dependencies, the import chain between two modules, its architectural layer, and import cycles. This lets the agent read repository organization straight from the import graph rather than reconstructing it through many searches.

\paragraph{Docs} The documentation view captures what the other three cannot: the \emph{intent} behind the code and the \emph{vocabulary} of its problem domain, drawn from the docstrings, comments, and signatures attached to each symbol. This makes it the natural view for queries phrased in human terms rather than symbol names. Because documentation has no inherent graph structure, the docs server has no edges to walk: rather than expand a neighborhood and rank it by centrality, it ranks documented-symbol nodes directly by full-text relevance over their parsed doc comments and signatures. Its \texttt{explore}/\texttt{node} pair returns the symbols whose documentation best matches a natural-language query; further tools (Table~\ref{tab:reps}) give module- and repository-level documented overviews. The docs view thus lets a query in plain words find the code that implements it even when the agent does not know the exact symbol names.

\paragraph{Hybrid} The hybrid view is not a new structure but a \emph{fusion} of the others, combining them into a single map so the agent can reason over several relations at once. It merges the call and dependency graphs into one graph whose nodes are both functions and files and whose edges include \texttt{calls}, \texttt{imports}, and \texttt{contains} (file\,$\to$\,definition); the \texttt{contains} edges carry the AST skeleton, so a single map holds containment, calls, and module dependencies together. Its \texttt{explore}/\texttt{node} pair ranks code by call and import connectivity jointly and returns a symbol's combined neighborhood; further tools (Table~\ref{tab:reps}) estimate a change's transitive blast radius and give a documented repository overview.

\subsection{Implementation of Code Representations}
\label{sec:approach}

%\song{this section reads like an experiment setting, not an approach}

We study how different code representations affect agent behavior by exposing each representation as a dedicated Model Context Protocol (MCP)\cite{anthropic2024mcp} tool server. An agent run is parameterized by a tuple:
\[
\langle \text{agent}, \text{model}, \text{representation}, \text{benchmark task}\rangle.
\]
Here, \textit{agent} denotes the coding-agent framework and \textit{model} denotes the LLM snapshot used by that framework. The representation determines which MCP server, if any, is mounted, and the benchmark task supplies the input prompt and scoring signal. When comparing representations, we keep the agent framework, model, and benchmark task fixed.

We implement five representation servers: AST, call graph, dependency graph, documentation, and Hybrid, which combines the other four. Before each agent run, the selected server indexes the target repository and stores the extracted information as a searchable repository map. All servers parse source files with tree-sitter\footnote{\url{https://tree-sitter.github.io/tree-sitter/}} using the appropriate grammar for each language. For \textbf{Python}, call relations are resolved with jedi\footnote{\url{https://github.com/davidhalter/jedi}}, import dependencies are extracted from \texttt{import} and \texttt{from} statements, and docstrings are loaded with griffe\footnote{\url{https://github.com/mkdocstrings/griffe}}. For \textbf{C/C++} repositories, where jedi and griffe do not apply, calls are resolved by matching call names to repository-level definitions, dependencies are extracted from \texttt{\#include} directives, and documentation is taken from comment blocks immediately preceding each definition.

All five controlled servers expose the same core navigation pair, \texttt{<rep>\_explore} and \texttt{<rep>\_node}, plus representation-specific utilities. This gives the agent a comparable entry point across workflows, while still allowing each representation to expose relation-specific operations. A map is a graph whose nodes denote code units, such as functions, files, or documented symbols, and whose edges encode the relation used by that representation, such as file-to-definition containment, calls, or imports.

All servers use the same basic query process. First, the server finds relevant seed nodes using full-text search over an SQLite FTS5 index ranked with BM25 \cite{robertson2009probabilistic}. Matches in symbol names are weighted more heavily than matches in signatures, documentation, or code bodies. For the structural servers, these seed nodes are then expanded through the server's graph relations: containment for AST, call edges for call graph, import edges for dependencies, and all merged edge types for Hybrid. The expanded graph neighborhood is ranked with random-walk-with-restart centrality \cite{tong2006fast}, starting from the seed nodes, and the highest-ranked source snippets are returned grouped by file. The documentation server is different because it has no structural edges: it skips graph expansion and directly returns documented symbols ranked by BM25. This keeps the retrieval entry point comparable across servers while varying the main information each map provides. Table~\ref{tab:reps} summarizes the tools, and Figure~\ref{fig:representation-listings} shows how the same code fragment appears under each representation.

\begin{figure}[t]
\centering
\begin{lstlisting}[basicstyle=\ttfamily\scriptsize,breaklines=true,escapechar=|]
|\textbf{(a) Source excerpt returned in the trajectory}|
|\hspace{0.6cm}\textbf{technical\_404\_response  (django/views/debug.py:459)}|
def technical_404_response(request, exception):
    """Create a technical 404 error response. `exception` is the Http404."""
    ...
    try:
        resolver_match = resolve(request.path)
    except Resolver404:
        pass

|\textbf{(b) Representation excerpts around the same source}|
# [AST] ast_explore:
  contains: debug.py -> debug
  contains: debug -> technical_404_response

# [CG] callgraph_explore:
  calls:    response_for_exception -> technical_404_response

# [Deps] deps_explore:
  imports:  test_debug.py -> debug.py
  imports:  debug.py -> urls.py

# [Docs] docs_explore:
  doc:      /** Create a technical 404 error response. `exception` is the Http404. */

# [Hybrid] hybrid_explore:
  calls:    get_object -> to_python
  source:   // converters.py  (django/urls/converters.py:1)
  source:   // technical_404_response  (django/views/debug.py:459)
\end{lstlisting}
% \caption{Abridged MCP response excerpts shown to Claude for the SWT-Bench task \texttt{django\_\_django-11620}. Panel (a) shows the real source excerpt; panel (b) shows the representation-specific relation or documentation lines returned around the same task.}
\vspace{-0.1in}
\caption{Abridged source and representation excerpts from \texttt{django\_\_django-11620}.}
\label{fig:representation-listings}
\end{figure}

\section{Experimental Setup}
\label{sec:exp}

%\song{the RQ1 does not make sense, please update all the RQs; overall, you want to explore the effectiveness of different types of code representation for different tasks; keep this in mind}
%\jiho{Addressed by framing the RQs around generic representation effectiveness, isolated representation effects, and trajectory analysis}
\subsection{Research Questions}
\label{sec:exp:rq}

\begin{itemize}
    \item \textbf{RQ1 (Effectiveness and cost of representation workflows).}
How do map-based code-representation workflows affect task success, token usage, wall time, and native tool use compared with a no-MCP baseline?
    \item \textbf{RQ2 (Trajectory-level strengths and limitations).}
When do representations help or hurt, and what trajectory patterns explain differences across task families?
\end{itemize}

RQ1 measures the aggregate behavior of each representation workflow relative to the baseline.
RQ2 examines trajectories where the representation and baseline differ to identify when a representation is beneficial, adds overhead, or fails to affect the decisive patch or test-generation step.

\begin{table*}[t!]
\centering
\footnotesize
\renewcommand{\arraystretch}{1.05}
\caption{Benchmarks and task-success metrics used in our evaluation.}
\label{tab:benchmarks-metrics}

\resizebox{\textwidth}{!}{%
\begin{tabular}{p{2.4cm}p{3.0cm}p{2.8cm}p{7.8cm}c}
\toprule
\textbf{Benchmark} & \textbf{Task} & \textbf{Agent output} & \textbf{Task-success metric} & \textbf{Sample} \\
\midrule
\textbf{LinuxFLBench} \cite{zhou2026taming} & Fault localization from a Linux kernel bug report. & Ranked list of up to ten likely buggy files. & File-level recall@$k$, for $k \in \{1,10\}$; success means the ground-truth file appears in the top-$k$ predictions. & 152 \\
\midrule
\textbf{SWE-bench Pro} \cite{deng2025swe} & Program repair from a real GitHub issue. & A code patch in unified-diff format. & Resolution rate; a task is resolved if the patch passes all hidden fail-to-pass tests and preserves all pass-to-pass tests. & 253 \\
\midrule
\textbf{SWT-bench} \cite{mundler2024swt} & Test generation from a GitHub issue. & A unit test that reproduces the issue. & Reproducing-test rate; the test must fail on the buggy version and pass after the known fix. & 161 \\
\midrule
\textbf{SecBench} \cite{lee2026sec} & Automatic security patching for vulnerabilities & A security patch. & Patch-resolution rate; a task is resolved if the generated patch is accepted by the benchmark harness. & 166 \\
\bottomrule
\end{tabular}%
}
\end{table*}
%This is a strong baseline because the agent's default tools already include the same \texttt{Glob} and \texttt{Grep} that the representation servers partly subsume, and because the underlying model can already reason about code at a high level without explicit structural input.

%\song{add the CodeGraph?}

\subsection{Experimental Grid}
\label{sec:approach:grid}

We evaluate a controlled grid of \emph{workflow} $\times$ \emph{benchmark}, using the same sampled task IDs across workflows. The controlled workflows are the no-MCP baseline and five representation workflows: AST, callgraph, deps, docs, and hybrid. Thus, the controlled grid contains:
\[
\underbrace{(1\,(\text{baseline}) + 5\,(\text{representation workflows}))}_{\text{6 controlled workflows per benchmark}} \times 4\,\text{benchmarks}.
\]

The baseline does not use any MCP-based representation tool. The agent only has access to its default repository interaction tools, including file reading and writing, glob search, grep search, and shell execution. As a result, it must infer repository structure, such as symbol definitions, call relationships, and file dependencies, through ordinary source-code exploration.

\subsection{Agent Stacks}
\label{sec:approach:agents}

We run experiments with \textbf{Claude Code} \cite{Claude} using \texttt{claude-haiku-4-5} at \texttt{effort=low}; the provider resolves this alias to \texttt{claude-haiku-4-5-20251001}. We access the model through the Anthropic API and Claude Code CLI's MCP integration. We pin built-in tools to \texttt{Read}, \texttt{Edit}, \texttt{Write}, \texttt{Bash}, \texttt{Grep}, and \texttt{Glob} via \texttt{--tools}. Representation workflows attach the relevant stdio MCP server through \texttt{--mcp-config}, making its tools available alongside built-ins, whereas the baseline attaches none.

The MCP-server design has two properties that matter for the study.
\emph{Agent-agnostic:} the same representation server can be evaluated on Claude Code, Codex~\cite{Codex}, or any future MCP-compatible host without modification.
\emph{Agent-executed retrieval:} we do not pre-compute and paste a fixed representation into the prompt; the agent issues representation queries itself and integrates the returned views into its trajectory.
The baseline keeps the same model, effort setting, and benchmark prompt, but the non-baseline treatment also includes the representation-use steer described below.

\begin{figure*}[t!]
  \centering
  \begin{subfigure}[t]{0.48\textwidth}
    \centering
    \includegraphics[width=0.9\linewidth]{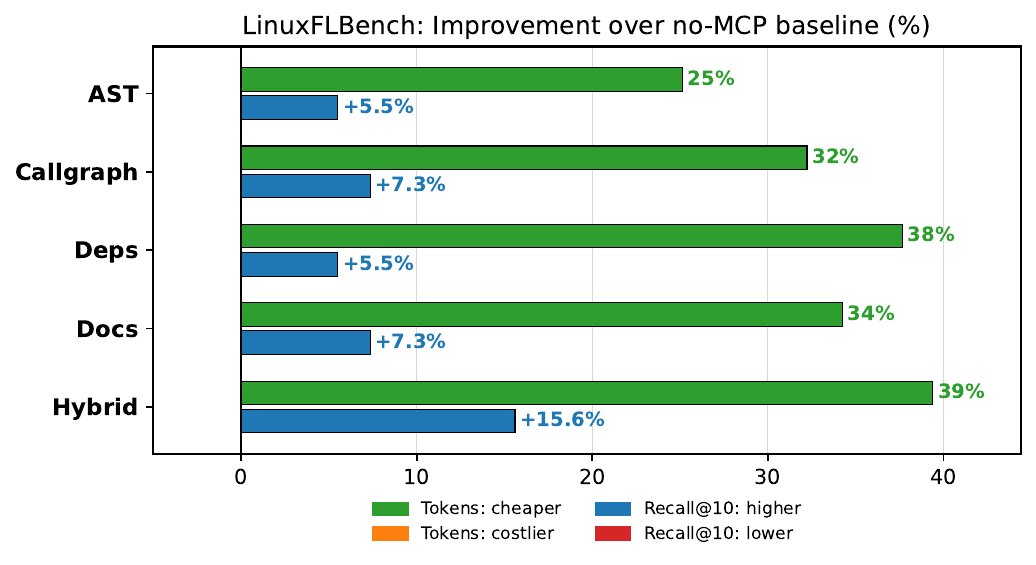}
    \caption{\textbf{LinuxFLBench} (fault localization)}
    \label{fig:linuxfl-relbar}
  \end{subfigure}
  \hfill
  \begin{subfigure}[t]{0.48\textwidth}
    \centering
    \includegraphics[width=0.9\linewidth]{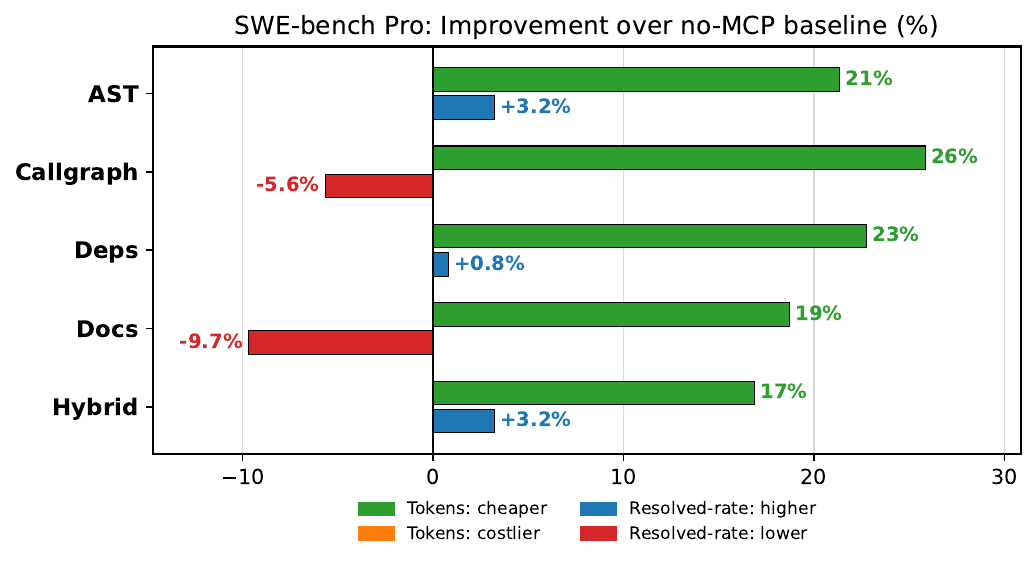}
    \caption{\textbf{SWE-bench Pro} (program repair)}
    \label{fig:swebench-relbar}
  \end{subfigure}

  \smallskip

  \begin{subfigure}[t]{0.48\textwidth}
    \centering
    \includegraphics[width=0.9\linewidth]{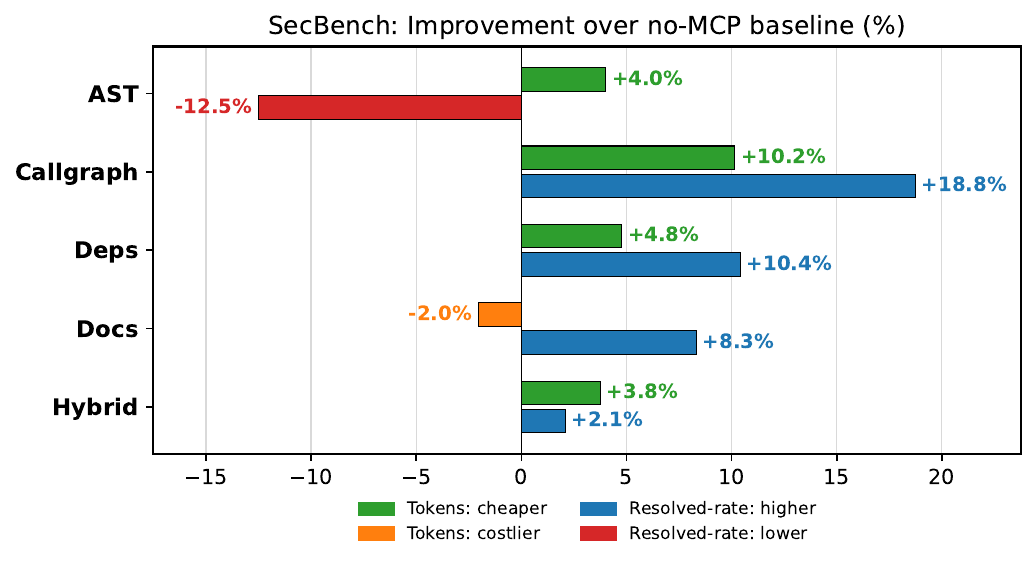}
    \caption{\textbf{SecBench} (security patching)}
    \label{fig:secbench-relbar}
  \end{subfigure}
  \hfill
  \begin{subfigure}[t]{0.48\textwidth}
    \centering
    \includegraphics[width=0.9\linewidth]{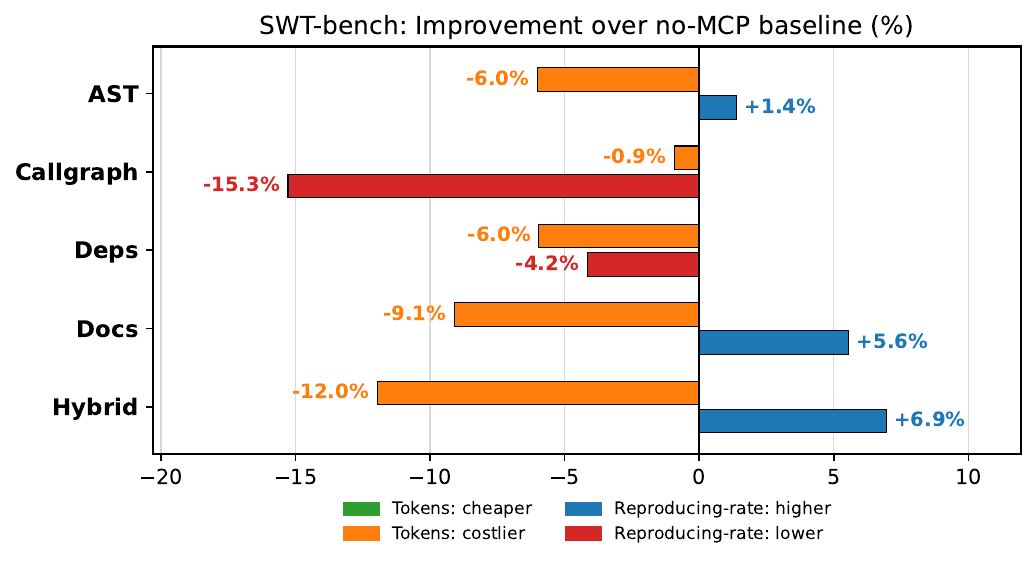}
    \caption{\textbf{SWT-bench} (test generation)}
    \label{fig:swtbench-relbar}
  \end{subfigure}
  \caption{Each representation's percent change vs.\ the no-MCP baseline on both axes (accuracy and tokens), across the four tasks. Green/blue mark improvements (cheaper/more accurate); orange/red mark regressions.}
  \label{fig:all-relbar}
\end{figure*}

\subsection{Benchmarks and Metrics}
\label{sec:exp:benchmarks-metrics}

We evaluate four benchmarks covering fault localization, program repair, test generation, and security patching. Table~\ref{tab:benchmarks-metrics} summarizes the task, expected agent output, success metric, and sample size for each benchmark.

%For LinuxFLBench, the agent is explicitly asked to output a JSON list of up to ten candidate files ordered from most to least likely; we parse this list to compute recall@$k$.

\paragraph{Sampling} We sampled each benchmark's task pool at the 95\% confidence level with a 5\% margin of error from a population proportion estimated by a pre-study pilot, following standard sample-size estimation practice~\cite{cochran_sampling}. The resulting sample sizes preserve representativeness while keeping the experimental budget tractable across all workflows. Samples are deterministic given a random seed; the same task IDs are used for every workflow.

\paragraph{Behavior and Cost Metrics.}
We report per-task averages for token usage, wall-clock time, MCP calls, and native tool calls. We use total token usage as our cost measure. The result tables report \texttt{avg\_wall\_s} as end-to-end wall-clock time in seconds, \texttt{avg\_MCP} as calls to MCP representation tools, and \texttt{avg\_builtin} as calls to the agent's native tools. We further break native tool use into \texttt{avg\_read}, \texttt{avg\_grep}, \texttt{avg\_glob}, and \texttt{avg\_bash}; \texttt{avg\_builtin} also includes native edit/write calls when the task uses them.

\section{Results}
\label{sec:results}

We organize the results to answer the two research questions we asked in Section~\ref{sec:exp:rq}. %RQ1 (Section~\ref{sec:results:rq1}) asks whether code representations help at all: for each benchmark, we compare the five representations against the no-MCP baseline, looking at both task success and efficiency (tokens, time, and tool usage). RQ2 (Section~\ref{sec:results:rq2}) then turns to the agent trajectories to explain why the representations differ, relating each representation's strengths and limitations to the structural relations it exposes.

\subsection{RQ1: Effectiveness and Cost of Representation Workflows}
\label{sec:results:rq1}

% RQ1 asks whether giving the agent a structured code representation, delivered as an MCP map, improves task success and efficiency relative to the no-MCP baseline, which sees the repository only through standard file and shell tools. We answer it per benchmark, comparing each of the five representations (AST, Callgraph, Deps, Docs, Hybrid) against the baseline on both the task-success metric and the behavior/cost envelope (tokens, wall time, cost, and tool usage). All conditions are run on the same instances.

RQ1 examines how representation workflows affect task success and efficiency compared with the no-MCP baseline. We report the results per benchmark because the effect depends strongly on where the task bottleneck lies: localization, patch synthesis, security reasoning, or test authoring.

\paragraph{Fault Localization (LinuxFLBench)}
Table~\ref{tab:linuxfl} reports file-localization recall and the behavior/cost metrics, and Figure~\ref{fig:all-relbar}(\subref{fig:linuxfl-relbar}) shows each representation's change relative to the baseline on performance and token cost axes. On LinuxFLBench, the code representations provide the clearest benefit. All five improve recall@10 over the baseline (0.717). Hybrid is highest at 0.829, Docs and Callgraph reach 0.770, and AST and Deps reach 0.757. At the same time, every representation cuts average token usage by 25--39\%, with Hybrid the cheapest ($-39\%$). The recall@1 differences are smaller and mixed, which suggests that the main benefit is improving the top-10 candidate set rather than reliably producing the first-ranked file. The mechanism behind the token savings is visible in the tool-usage columns: the baseline assembles its answer through 8 reads, 6 greps, and 10 shell calls per task, whereas the representation agents replace almost all of that browsing with 9--13 map calls and drop native \texttt{Read} to at most 1.1 and \texttt{Grep}/\texttt{Glob} to essentially zero. %\song {missing details, you need to give the results of each of the five representations compared to the baseline}
% Here, representations help on both dimensions at once. Every representation matches or improves file recall@10 over the baseline (0.717). The representations all reach 0.757--0.829, and the hybrid representation is the strongest localizer at 0.829, a $+16\%$ relative gain. The recall@1 differences are small and mixed (0.529--0.586 versus the baseline's 0.571), so the improvement is concentrated in the broader top-10 candidate set rather than in pinpoint first-guess accuracy. At the same time, every representation is substantially cheaper:
% total token usage falls by 25--39\% (from 1.01M for the baseline to 0.61--0.75M), with the hybrid representation both the most accurate and the cheapest (0.61M
% tokens, $39\%$ fewer than the baseline). The mechanism behind the token savings is visible in the tool-usage columns: the baseline assembles its answer through
% roughly 8 reads, 6 greps, and 10 shell calls per task, whereas the representation agents replace almost all of that browsing with 9--13 map calls and drop native \texttt{Read} to under 1.1 and \texttt{Grep}/\texttt{Glob} to essentially zero. The one cost the representations pay is wall-clock time: the additional MCP round-trips raise end-to-end latency from 67s to 82--106s, even though they reduce tokens. In short, on fault localization, every representation dominates the baseline on the accuracy--cost plane, and the hybrid representation is the clear winner on both.

\begin{table*}[t]
\centering
\footnotesize
\setlength{\tabcolsep}{7pt}
\renewcommand{\arraystretch}{0.9}
\caption{LinuxFLBench: file-localization recall and behavior/cost metrics per workflow}
\label{tab:linuxfl}
\resizebox{1\textwidth}{!}{%
\begin{tabular}{l rr rr rr rrrr}
\toprule
Workflow & fR@1 & fR@10 & avg\_tokens & avg\_wall\_s & avg\_MCP & avg\_builtin & avg\_read & avg\_grep & avg\_glob & avg\_bash \\
\midrule
Baseline  & 0.572 & 0.717 & 1,008,163 & 66.7  & 0.0  & 25.4 & 8.4 & 6.0 & 1.4 & 9.5 \\ \hline
AST       & 0.526 & 0.757 & 754,768   & 104.6 & 13.1 & 0.9  & 0.5 & 0.0 & 0.0 & 0.3 \\
Callgraph & 0.572 & 0.770 & 683,227   & 105.1 & 11.0 & 2.2  & 1.1 & 0.0 & 0.0 & 1.1 \\
Deps      & 0.572 & 0.757 & 628,244   & 87.6  & 8.5  & 0.7  & 0.4 & 0.0 & 0.0 & 0.3 \\
Docs      & \textbf{0.586} & 0.770 & 662,834   & 105.6 & 11.1 & 2.4  & 0.6 & 0.0 & 0.0 & 1.7 \\
Hybrid    & 0.526 & \textbf{0.829} & \textbf{610,910}   & 82.2  & 9.5  & 0.5  & 0.3 & 0.0 & 0.0 & 0.3 \\
\bottomrule
\end{tabular}}
\end{table*}

\paragraph{Program repair (SWE-bench Pro)}
Table~\ref{tab:swebench} reports the resolution rate and behavior/cost metrics, and Figure~\ref{fig:all-relbar}(\subref{fig:swebench-relbar}) shows the change relative to the baseline.
On SWE-bench Pro, the efficiency win persists, but the accuracy effect largely vanishes. Every representation reduces token usage by 17--26\% relative to the baseline (3.12M tokens), with Callgraph the cheapest (2.31M, $-26\%$) and the hybrid the least aggressive (2.59M, $-17\%$); cost and native tool usage move in the same direction, with \texttt{Read} falling from 12.7 to 7.6--8.7 per task. The resolution rate, however, stays close to the baseline (49.0\%): AST and Hybrid reach 50.6\% and Deps 49.4\% (gains within noise), while Callgraph (46.2\%) and Docs (44.3\%) fall slightly below it. Wall-clock time decreases (269s for the baseline versus 194--209s for the representations), because the agent spends fewer turns browsing. %In conclusion, on a demanding, edit-and-test repair task, the representations preserve task success while making the agent meaningfully cheaper and faster. The benefit is efficiency at equal accuracy, not higher accuracy.
%On SWE-bench Pro, the main benefit is efficiency rather than resolution. The representations reduce token usage by 17--26\% and reduce wall time relative to the baseline, while resolution remains close to the baseline. Because each instance is run once, the small resolution differences should be interpreted as descriptive rather than stable accuracy gains. \song {missing details, you need to give the results of each of the five representations compared to the baseline}

\begin{table*}[t]
\centering
\footnotesize
\setlength{\tabcolsep}{7pt}
\renewcommand{\arraystretch}{0.9}
\caption{SWE-bench Pro: patch-resolution rate and behavior/cost metrics per workflow.}
\label{tab:swebench}
\resizebox{1\textwidth}{!}{%
\begin{tabular}{l r rr rr rrrr}
\toprule
Workflow & Resolved & avg\_tokens & avg\_wall\_s & avg\_MCP & avg\_builtin & avg\_read & avg\_grep & avg\_glob & avg\_bash \\
\midrule
Baseline  & 49.0\% & 3,116,193 & 268.8 & 0.0 & 44.5 & 12.7 & 0.0 & 0.0 & 25.2 \\ \hline
AST       & \textbf{50.6\%} & 2,450,752 & 209.1 & 5.2 & 37.7 & 8.4 & 0.2 & 0.0 & 23.4 \\
Callgraph & 46.2\% & \textbf{2,310,653} & 193.7 & 4.2 & 38.1 & 8.7 & 0.2 & 0.1 & 23.6 \\
Deps      & 49.4\% & 2,406,690 & 205.3 & 3.4 & 39.1 & 8.6 & 0.2 & 0.0 & 24.7 \\
Docs      & 44.3\% & 2,532,738 & 205.8 & 4.9 & 39.0 & 7.6 & 0.1 & 0.1 & 25.8 \\
Hybrid    & \textbf{50.6\%} & 2,590,046 & 204.3 & 4.3 & 39.2 & 8.6 & 0.2 & 0.0 & 24.5 \\
\bottomrule
\end{tabular}}
\end{table*}

\paragraph{Test Generation (SWT-bench)}
Table~\ref{tab:swtbench} reports the reproducing-test rate and behavior/cost metrics, and Figure~\ref{fig:all-relbar}(\subref{fig:swtbench-relbar}) shows the change relative to the baseline.
% On SWT-bench, success means producing a reproducing test: a generated test that fails on the buggy program and passes after the gold fix. The no-MCP baseline solves $72/161$ instances ($44.7\%$). The representation results are mixed but mostly close to the baseline. Hybrid is the strongest condition at $77/161$ ($47.8\%$, $+3.1$ percentage points), followed by Docs at $76/161$ ($47.2\%$, $+2.5$ points), CodeGraph at $75/161$ ($46.6\%$, $+1.9$ points), and AST at $73/161$ ($45.3\%$, $+0.6$ points). Deps falls slightly below the baseline ($69/161$, $42.9\%$), and Callgraph is substantially worse ($61/161$, $37.9\%$). Thus, for test generation, the best representations provide a small accuracy gain, but the gain is not as broad as in fault localization: tests require not only finding the relevant behavior, but also writing a discriminating assertion that exercises the bug.
On SWT-bench, the success differences are small. {Hybrid and Docs have the highest reproducing-test rates in this sample (47.8\% and 47.2\%), improving over the 44.7\% baseline by +3.1 and +2.5 percentage points. AST is nearly tied with the baseline (45.3\%, +0.6 points), while Deps falls slightly below it (42.9\%, -1.9 points) and Callgraph is substantially worse (37.9\%, -6.8 points). At the same time, all representation workflows use more tokens than the baseline.} We interpret SWT-bench as a case where representation tools can slightly change outcomes, but the decisive step is often writing the right assertion rather than finding the right code. {The behavior profile is different from LinuxFLBench and SWE-bench Pro: every representation reduces native shell/file-tool usage, but the added map context raises token use and wall time for all workflows. Wall time increases from 161s for the baseline to 170--201s for the representation workflows. Native/MCP tool-call averages are 27.3/5.7 for AST, 31.9/3.5 for Callgraph, 30.2/2.6 for Deps, 30.8/4.2 for Docs, and 28.6/3.8 for Hybrid, compared with 33.4/0.0 for the baseline.}
%\song {missing details, you need to give the results of each of the five representations compared to the baseline}
%The baseline uses 33.4 native tool calls per task on average, while the representation conditions use 27.3--31.9 native calls and add 2.6--5.7 MCP calls. Wall time is mixed: CodeGraph is faster than the baseline (139s versus 161s), but the controlled representation workflows take 170--201s. In short, SWT-bench shows a modest effectiveness benefit for Docs and Hybrid, but not an efficiency benefit; the maps trade fewer native searches for more structured context and higher token use.
% The cost profile is also different from LinuxFLBench and SWE-bench Pro. Although every representation reduces native shell/file-tool usage, none reduces API cost. The baseline uses $33.4$ native tool calls per task on average, while the representation conditions use $17.1$--$31.9$ native calls and add $2.6$--$5.7$ MCP calls. The extra context from those MCP calls raises average cost from $\$0.312$ for the baseline to $\$0.375$--$\$0.421$ ($+20$--$35\%$). Wall time is mixed: CodeGraph is faster than the baseline (139s versus 161s), but the per-representation MCP conditions take 170--201s. In short, SWT-bench shows a modest effectiveness benefit for Docs and Hybrid, but not an efficiency benefit; the representations trade fewer native searches for more structured context and higher cost.

\begin{table*}[t]
\centering
\footnotesize
\setlength{\tabcolsep}{7pt}
\renewcommand{\arraystretch}{0.9}
\caption{SWT-bench: reproducing-test rate and behavior/cost metrics per workflow.}
\label{tab:swtbench}
\resizebox{1\textwidth}{!}{%
\begin{tabular}{l r rr rr rrrr}
\toprule
Workflow & Reprod. & avg\_tokens & avg\_wall\_s & avg\_MCP & avg\_builtin & avg\_read & avg\_grep & avg\_glob & avg\_bash \\
\midrule
Baseline  & 44.7\% & \textbf{1,671,772} & 161.4 & 0.0 & 33.4 & 12.7 & 0.0 & 0.0 & 16.9 \\ \hline
AST       & 45.3\% & 1,772,180 & 170.1 & 5.7 & 27.3 & 9.7  & 0.0 & 0.0 & 13.6 \\
Callgraph & 37.9\% & 1,687,094 & 201.1 & 3.5 & 31.9 & 12.1 & 0.0 & 0.0 & 15.8 \\
Deps      & 42.9\% & 1,771,300 & 184.2 & 2.6 & 30.2 & 10.4 & 0.0 & 0.0 & 15.7 \\
Docs      & 47.2\% & 1,823,938 & 185.1 & 4.2 & 30.8 & 10.9 & 0.0 & 0.0 & 16.1 \\
Hybrid    & \textbf{47.8\%} & 1,871,848 & 192.5 & 3.8 & 28.6 & 10.5 & 0.0 & 0.0 & 14.2 \\
\bottomrule
\end{tabular}}
\end{table*}

\paragraph{Security Patch Evaluation (SecBench)}
Table~\ref{tab:secbench} reports the security-patch resolution rate and behavior/cost metrics, and Figure~\ref{fig:all-relbar}(\subref{fig:secbench-relbar}) shows the change relative to the baseline.
% On SecBench, the success metric is whether the generated security patch is evaluated as resolved by the benchmark harness. The baseline resolves $48/166$ instances ($28.9\%$). Here, the best single relation is Callgraph: it resolves $57/166$ instances ($34.3\%$), a $+5.4$ percentage-point gain over the baseline. CodeGraph ($54/166$, $32.5\%$), Deps ($53/166$, $31.9\%$), and Docs ($52/166$, $31.3\%$) also improve over the baseline, while Hybrid is nearly tied ($49/166$, $29.5\%$). AST is the only representation that hurts accuracy, dropping to $42/166$ ($25.3\%$). This pattern is consistent with the task shape: security patches often require following a crash or sanitizer report through callers, callees, and dependency context, so relation-level navigation helps; a pure containment view is less reliable.
On SecBench, {Callgraph has the highest controlled patch-resolution rate in this sample (34.3\%, +5.4 points over the 28.9\% baseline). Deps (31.9\%, +3.0 points) and Docs (31.3\%, +2.4 points) also improve resolution; Hybrid is nearly tied (29.5\%, +0.6 points), and AST drops below the baseline (25.3\%, -3.6 points).} The efficiency picture is mixed. {In token use, AST (656K), Callgraph (614K), Deps (651K), and Hybrid (658K) are below the 684K baseline, while Docs is slightly above it (698K). Wall time increases for every representation workflow, from 136s for Hybrid to 290s for Docs, compared with 66s for the baseline.} They also sharply reduce native file/shell browsing: the baseline uses 12.9 native tool calls per task, while the representation workflows use 6.5--8.0 native calls plus 3.1--5.5 MCP calls. {By workflow, the native/MCP tool-call averages are 6.5/5.5 for AST, 7.4/3.1 for Callgraph, 7.6/3.1 for Deps, 7.4/4.4 for Docs, and 8.0/5.0 for Hybrid, compared with 12.9/0.0 for the baseline.} SecBench therefore suggests that execution-path information can improve patch resolution while shifting work from native browsing to map queries and increasing latency.
%\song{missing details, you need to give the results of each of the five representations compared to the baseline}
% The efficiency trade-off is again unfavorable. The baseline averages $\$0.164$ per task and 66s wall time. All representation conditions cost more, at $\$0.207$--$\$0.237$ per task ($+26$--$44\%$), and all take longer, from 136s for Hybrid to 290s for Docs. They do, however, sharply reduce native file/shell browsing: the baseline uses $12.9$ native tool calls per task, while the representation conditions use $4.7$--$8.0$ native calls plus $3.1$--$5.5$ MCP calls. SecBench therefore shows the clearest accuracy gain for Callgraph, with smaller gains for CodeGraph, Deps, and Docs, but those gains come with higher cost and latency rather than savings.

\begin{table*}[t]
\centering
\footnotesize
\setlength{\tabcolsep}{7pt}
\renewcommand{\arraystretch}{0.9}
\caption{SecBench: security-patch resolution rate and behavior/cost metrics per workflow.}
\label{tab:secbench}
\resizebox{1\textwidth}{!}{%
\begin{tabular}{l r rr rr rrrr}
\toprule
Workflow & Resolved & avg\_tokens & avg\_wall\_s & avg\_MCP & avg\_builtin & avg\_read & avg\_grep & avg\_glob & avg\_bash \\
\midrule
Baseline  & 28.9\% & 683,774 & 66.4  & 0.0 & 12.9 & 7.2 & 0.0 & 0.0 & 4.1 \\ \hline
AST       & 25.3\% & 656,332 & 166.6 & 5.5 & 6.5  & 3.6 & 0.0 & 0.0 & 1.3 \\
Callgraph & \textbf{34.3\%} & \textbf{614,360} & 184.2 & 3.1 & 7.4  & 4.1 & 0.0 & 0.0 & 1.6 \\
Deps      & 31.9\% & 651,116 & 144.4 & 3.1 & 7.6  & 4.4 & 0.0 & 0.0 & 1.5 \\
Docs      & 31.3\% & 697,763 & 290.4 & 4.4 & 7.4  & 4.3 & 0.0 & 0.0 & 1.5 \\
Hybrid    & 29.5\% & 658,069 & 136.4 & 5.0 & 8.0  & 4.7 & 0.0 & 0.0 & 1.1 \\
\bottomrule
\end{tabular}}
\end{table*}

\begin{findingbox}
\textbf{RQ1 takeaway.}
%\song{update here, given an overall message, and then details of each}
Code representations are primarily an efficiency tool, and their payoff is set by where the task's bottleneck lies: they help most when the hard problem is finding code and fade when it is writing the fix or test. On LinuxFLBench, they improve recall@10 and reduce token use, and on SWE-bench Pro they mainly reduce token cost. On SecBench, they provide some success gains, especially with the call graph, but efficiency remains close to the baseline with only small token reductions. On SWT-bench, success gains are smaller and come with higher token use. %Overall, the benefit of each representation depends on the task bottleneck.
\end{findingbox}

\subsection{RQ2: Qualitative Trajectory Analysis}
\label{sec:results:rq2}

%RQ1 shows whether representations help overall, but aggregate metrics do not explain why they help or hurt. 
%In this RQ, we analyze the agent trajectories for cases where a representation differs from the baseline in success or token use. 

%\song{I do not understand why you need these arbitrary settings/definitions in a qualitative analysis. Please keep this simple, just show compared to baselines: 1. how many more/less tasks in a benchmark a representation resolved; 2. how many tasks in a benchmark a representation consumed fewer/more tokens, 3. how many tasks cost less/more,}

In this RQ, we analyze the agent trajectories for the instances where a representation diverges from the baseline. We compare each representation to the baseline on two axes. On the \textbf{performance} axis, we count, per benchmark, how many instances the representation \emph{wins} (solves where the baseline fails) and how many it \emph{loses} (fails where the baseline solves). On the \textbf{token} axis, we count how many instances the representation runs on \emph{fewer} tokens than the baseline and how many on \emph{more}. These counts, summarized in Table~\ref{tab:rq2-landscape}, reveal the main strengths and weaknesses of each representation and how they depend on the task family; we then read the underlying trajectories to explain why each divergence occurs.

\subsubsection{The landscape of divergences}
%Table~\ref{tab:rq2-landscape} summarizes where each representation diverges from the baseline in either task success or token use. Performance counts rescues, regressions, and their net difference; Avg and Med report the mean and median token ratio, and bold marks the best value within each benchmark block. The main pattern follows the task bottleneck. On \textbf{LinuxFLBench}, where the task is pure localization, representations improve both axes: they produce more rescues than regressions and usually reduce token use. On \textbf{SWE-bench Pro}, performance is roughly balanced, but token savings remain strong, suggesting that maps mostly replace repository browsing rather than directly improving patch quality. On \textbf{SecBench}, the call graph gives the clearest success gain, while efficiency stays close to the baseline with only small average token savings. On \textbf{SWT-bench}, success gains are weaker, and most representations add token overhead. Overall, representations help most when finding the relevant code is the bottleneck; their benefit shrinks when the hard part is writing the final patch or test. \song{update Table VII, rewrite the results in this part}

Table~\ref{tab:rq2-landscape} summarizes where each representation diverges from the baseline in either task success or token use, and the main pattern follows the task bottleneck. On \textbf{LinuxFLBench}, where the task is pure localization, representations win on both axes: every representation wins more often than it loses (Hybrid never loses, net $+17$), and roughly three quarters of the instances use fewer tokens than the baseline. On \textbf{SWE-bench Pro}, performance is roughly balanced (nets between $-12$ and $+4$), but the token advantage remains strong, with about three quarters of instances using fewer tokens, suggesting that maps mostly replace repository browsing rather than directly improving patch quality. On \textbf{SecBench}, the call graph gives the clearest success gain (net $+9$), but the token axis flips: for every representation, more instances use more tokens than fewer. On \textbf{SWT-bench}, success gains are weaker and the token axis more costly (only the call graph uses fewer tokens on a majority of instances). %Overall, representations help most when finding the relevant code is the bottleneck; their benefit shrinks, and their token cost grows, when the hard part is writing the final patch or test.

\begin{table}[t]
\centering
\caption{RQ2 divergence landscape. \textbf{Performance} counts, per benchmark, how many instances the representation \emph{wins} (solves where the baseline fails) versus \emph{loses} (the reverse), with their net. \textbf{Tokens} counts how many instances the representation runs on \emph{fewer} tokens than the baseline versus \emph{more}. Bold marks the best value within each benchmark block.}
\label{tab:rq2-landscape}
{\footnotesize
\renewcommand{\arraystretch}{0.85}
\setlength{\tabcolsep}{5pt}
\begin{tabular}{ll rrr rr}
\toprule
& & \multicolumn{3}{c}{Performance} & \multicolumn{2}{c}{Tokens} \\
\cmidrule(lr){3-5}\cmidrule(lr){6-7}
Benchmark & Rep & Win & Loss & Net & Fewer & More \\
\midrule
\multirow{5}{*}{\shortstack[l]{\textbf{Fault Loc.}\\(LinuxFLBench)}}
 & AST       & 17 & 11 & $+6$  & 111 & 41 \\
 & Callgraph & 17 & 9  & $+8$  & 109 & 43 \\
 & Deps      & 17 & 11 & $+6$  & 115 & 37 \\
 & Docs      & 17 & 9  & $+8$  & 115 & 37 \\
 & Hybrid    & 17 & \textbf{0} & \textbf{+17} & \textbf{117} & \textbf{35} \\
\midrule
\multirow{5}{*}{\shortstack[l]{\textbf{Program Repair}\\(SWE-bench Pro)}}
 & AST       & \textbf{28} & 24 & $+4$  & 190 & 63 \\
 & Callgraph & 20 & 27 & $-7$  & \textbf{197} & \textbf{56} \\
 & Deps      & 21 & 20 & $+1$  & 190 & 63 \\
 & Docs      & 21 & 33 & $-12$ & 181 & 72 \\
 & Hybrid    & 24 & 20 & \textbf{+4} & 181 & 72 \\
\midrule
\multirow{5}{*}{\shortstack[l]{\textbf{Security Patch.}\\(SecBench)}}
 & AST       & 9  & 15 & $-6$  & 75 & 91 \\
 & Callgraph & \textbf{18} & 9 & \textbf{+9} & \textbf{79} & \textbf{87} \\
 & Deps      & 14 & 9  & $+5$  & 78 & 88 \\
 & Docs      & 13 & 9  & $+4$  & 77 & 89 \\
 & Hybrid    & 13 & 12 & $+1$  & 74 & 92 \\
\midrule
\multirow{5}{*}{\shortstack[l]{\textbf{Test Gen.}\\(SWT-bench)}}
 & AST       & 27 & 26 & $+1$  & 73 & 88 \\
 & Callgraph & 18 & 29 & $-11$ & \textbf{85} & \textbf{76} \\
 & Deps      & 25 & 28 & $-3$  & 72 & 89 \\
 & Docs      & 25 & 21 & $+4$  & 59 & 102 \\
 & Hybrid    & \textbf{28} & 23 & \textbf{+5} & 67 & 94 \\
\bottomrule
\end{tabular}%
}
\end{table}

\subsubsection{Trajectory coding protocol}
%To explain the divergences in Table~\ref{tab:rq2-landscape}, we analyze paired baseline and representation trajectories that differ in task success or token use. For each pair, we inspect the returned map content, files read or edited, and final output, then assign the pair to the closest category in our codebook\song{what's your codebook?}, separately for the \textbf{performance} axis (rescues/regressions)\song{update this} and \textbf{token} axis (big-saves/costs)\song{update this}. The token buckets are interpretive rather than statistically derived: \textsc{Big-save} is $\le 0.66\times$ baseline tokens, \textsc{Save} is $0.66$--$0.90\times$, \textsc{Neutral} is $0.90$--$1.10\times$, and \textsc{Cost} is $>1.10\times$\song{update this, this is unrelated to your trajectory analysis}. Because bottlenecks shift from retrieval to editing or test authoring, we use benchmark-specific categories. They describe common trajectory patterns, not causal proof. If both runs reach the same files but differ only in patch or test quality, we code the case as one-shot patch/test variance. We report each count as $n/N$, where $N$ is the representation's total instances for that outcome on that benchmark. \song{n is what?}

%\song{this part is your protocol, but now in a chaos; you need to give detailed coding of how you are analyzing trajectory, open card? try to find a systematic approach}

The counts in Table~\ref{tab:rq2-landscape} say \emph{how often} a representation helps or hurts; to explain \emph{why}, we read the underlying trajectories. For every divergent instance (a win, a loss, or a run that uses fewer or more tokens than the baseline), two authors read the paired baseline and representation trajectories end to end, including the query issued, the map content returned, the files read or edited, and the final artifact, and write a short free-text reason for the divergence. We then group these reasons into recurring categories, iterating over the set until the categories are stable, and assign every divergent instance to one category. We code the performance axis and the token axis separately. Because the task bottleneck shifts across benchmarks, the recurring reasons genuinely differ per benchmark, so we derive a separate set of categories for each benchmark rather than forcing one shared vocabulary. The categories describe common trajectory patterns, not causal proof. Below we report each category as $n/N$, where $n$ is the number of instances assigned to that category and $N$ is the representation's total number of instances for that outcome (win, loss, fewer-token, or more-token) on that benchmark.

%To explain these divergences, we went past the counts and read every divergent trajectory end to end (the representation run and the baseline run on the same instance) to recover, for each one, the concrete reason the representation helped or hurt, on both the performance axis (why a rescue or a regression) and the token axis (why a big-save or a cost). Because the task's bottleneck moves from retrieval (localization) to editing and test-authoring, these reasons are genuinely different from one benchmark to the next, so we do not force a single shared vocabulary; we name the recurring reasons per benchmark and report each representation's most common one. Throughout, a category's instance count is given as $n/N$, where $N$ is that representation's total for the outcome (rescue, regression, big-save, or cost) on that benchmark. We analyze the two axes separately: the \textbf{performance} axis first and the \textbf{token} axis second.

\begin{table*}[t]
\centering
% \caption{RQ2 behavioral categories across the four study benchmarks. On \textbf{LinuxFLBench} (fault localization), each representation rescues and regresses on a \emph{different} bug shape---the shape its structural relation can or cannot reach. On the editing-bound benchmarks---\textbf{SWE-bench Pro} repair, \textbf{SecBench} security patching, and \textbf{SWT-bench} test generation---localization is less often the only bottleneck, so the dominant categories shift toward patch-invariant failures, one-shot patch/test variance, and additive map cost. Each cell gives the rank-1 category and its share ($N/M$) of that representation's rescues, regressions, big-saves, or costs.}
\caption{RQ2 behavioral categories. Each cell reports the top category and its share ($n/N$).}
\label{tab:rq2-dist-lflb-swe}
\resizebox{\textwidth}{!}{%
\begin{tabular}{l l l l l l}
\toprule
& & \multicolumn{2}{c}{\textbf{Performance axis}} & \multicolumn{2}{c}{\textbf{Token axis}} \\
\cmidrule(lr){3-4}\cmidrule(lr){5-6}
Benchmark & Rep & Dominant win & Dominant loss & Dominant fewer-token category & Dominant more-token category \\
\midrule
\textbf{Fault Localization} & AST       & Report-named in-file fault (17/17) & Symptom $\neq$ root-cause file (7/11) & Read displacement (79/111) & Map overhead (30/41) \\
(LinuxFLBench)          & Callgraph & Call-reachable cross-file fault (17/17) & Neighbor-drift (5/9) & Read displacement (109/109) & Map overhead (34/43) \\
                      & Deps      & Report names function/API (13/17) & Arch file outside import cluster (9/11) & Read displacement (72/115) & Map overhead (32/37) \\
                      & Docs      & Terms match in-file comments (17/17) & Symptom-only crash, no term (7/9) & Read displacement (93/115) & Over-exploration (27/37) \\
                      & Hybrid    & Lands the single target file (11/17) & --- (zero losses) & Read displacement (85/117) & Verbatim-source explosion (27/35) \\
\midrule
\textbf{Program Repair} & AST       & Report-named in-file fault (18/28) & Correct localization, wrong patch (16/24) & Read displacement (184/190) & Map overhead (58/63) \\
(SWE-bench Pro)       & Callgraph & Call-reachable cross-file fault (11/20) & Correct localization, wrong patch (22/27) & Read displacement (188/197) & Map overhead (35/56) \\
                       & Deps      & Single-module dump (17/21) & Correct localization, wrong patch (19/20) & Read displacement (186/190) & Generic/popular symbol name (51/63) \\
                       & Docs      & Cross-file shared-symbol bug (20/21) & Correct localization, wrong patch (23/33) & Read displacement (115/181) & Over-exploration (53/72) \\
                       & Hybrid    & Lands the single target file (14/24) & Correct localization, wrong patch (20/20) & Read displacement (174/181) & Verbatim-source explosion (44/72) \\
\midrule
\textbf{Security Patching} & AST       & Boundary guard (5/9) & Wrong/over-local sink (5/15) / Unmodeled invariant (5/15) & Abstention-cheapness (54/75) & Map overhead (55/91) \\
(SecBench)              & Callgraph & Flow localization (9/18) & Wrong/over-local sink (4/9) & Abstention-cheapness (51/79) & Map overhead (44/87) \\
                       & Deps      & Contract/API (6/14) & Wrong/over-local sink (4/9) & Abstention-cheapness (60/78) & Map overhead (44/88) \\
                       & Docs      & Contract/API (7/13) & Wrong/over-local sink (3/9) / Context dilution (3/9) & Abstention-cheapness (57/77) & Double-spend (47/89) \\
                       & Hybrid    & Flow localization (5/13) & Wrong/over-local sink (5/12) & Abstention-cheapness (51/74) & Map overhead (58/92) \\
\midrule
\textbf{Test Generation} & AST       & Test-variance win (25/27) & Correct localization, wrong patch/test (24/26) & Read displacement (73/73) & Map overhead (88/88) \\
(SWT-bench)            & Callgraph & Test-variance win (15/18) & Correct localization, wrong patch/test (27/29) & Read displacement (85/85) & Map overhead (76/76) \\
                       & Deps      & Test-variance win (22/25) & Correct localization, wrong patch/test (27/28) & Read displacement (72/72) & Map overhead (50/89) \\
                       & Docs      & Test-variance win (23/25) & Correct localization, wrong patch/test (19/21) & Read displacement (59/59) & Over-exploration (102/102) \\
                       & Hybrid    & Test-variance win (24/28) & Correct localization, wrong patch/test (22/23) & Read displacement (67/67) & Verbatim-source explosion (94/94) \\
\bottomrule
\end{tabular}}
\end{table*}

\subsubsection{Root Causes for Performance Benefit}

%\song{add the analysis of the four benchmarks}
Applying the protocol above to the performance axis yields the following categories, which we report per benchmark:

%\paragraph{Performance on fault localization}

%\song{I cannot understand how you get the categories? you make them up? What's the process of creating the categories?}

\paragraph{Performance on fault localization}
The first block of Table~\ref{tab:rq2-dist-lflb-swe} shows that fault-localization gains depend on whether a representation exposes the relation needed by the bug. For AST, all wins are \emph{report-named in-file faults}: the report names, or implies, the faulty function and the AST containment edge maps that symbol to its file ($17/17$). Its main loss category is \emph{symptom~$\neq$~root-cause file}: the report names a symptom, but the real fault is in a structurally distant sibling file that containment cannot reach ($7/11$). For callgraph, all wins are \emph{call-reachable cross-file faults}: the fix is in a caller or callee of the reported symbol, often in another file. Its losses are \emph{neighbor-drift}, where call edges stay near the wrong symptom file, and \emph{ranking dilution}, where the fault appears in the graph but is buried below highly connected helpers. For deps, most wins are \emph{report-names-API} faults, where the report names a function or call site and the dependency map retrieves the relevant file ($13/17$); its main loss is \emph{arch files outside the import cluster}, where architecture- or protocol-specific files are not import-reachable from the generic cluster ($9/11$). For docs, all wins are \emph{terms-match-in-comments}: report terms match comments or identifiers in the fault file. Its main loss is \emph{symptom-only crash, no term}, where the report gives only a panic/crash symptom and text search drifts to lexically similar wrong files ($7/9$). Finally, \emph{hybrid is the safe default}: it wins both report-named single-file faults ($11/17$) and cross-subsystem symptom-to-fault cases ($6/17$), and it has zero losses because it combines all relation types. Overall, single-relation maps each help a distinct bug shape and fail when that relation is missing, while Hybrid covers the widest range by combining all clues. Figure~\ref{fig:eg-callgraph} illustrates this match with a verified win/loss pair from the same representation, callgraph, contrasting it with the baseline on the same instance.

\vspace{-0.05in}
\rescueregress
{fig:eg-callgraph}
{A matched callgraph win/loss pair on LinuxFLBench%: the call relation crosses the symptom-to-fault layer when the fault is call-reachable (top), but drifts to the wrong neighbors when it is not (bottom).}
}
{Callgraph win: call-reachable cross-file fault (LinuxFLBench~14053)}
{\emph{Bug:} the kernel freezes during an rsync to an NFS mount; the fault is two layers below the symptom, in the RPC client \texttt{net/sunrpc/clnt.c}. Following the RPC call chain, callgraph pulls \texttt{clnt.c} into its map and ranks it \#9; AST and deps, whose relations stay within the \texttt{fs/nfs/*} symptom directory, never surface it, and the baseline misses it after 42 actions. \textbf{Only a call relation crosses the NFS$\rightarrow$RPC layer between symptom and fault.}}
{Callgraph loss: neighbor-drift off the call chain (LinuxFLBench~5093)}
{\emph{Bug:} a WiFi card hangs on suspend; the fault is in the driver \texttt{drivers/net/wireless/atmel.c}. Callgraph followed the firmware/resume call chains into PCMCIA glue and \texttt{drivers/base/power/resume.c} and \emph{never reached} \texttt{atmel.c}: the driver loads firmware through a path that is not a direct call edge, so it is absent from the map. The baseline's keyword search ranks \texttt{atmel.c} \#1. \textbf{When the fault is a sibling the symptom does not call into, the call map drifts to the wrong neighbors.}}

%\song{update rescures/regr globally, these two words missing leading}

\paragraph{Performance on program repair}

%\song{ again the same issue as above, I do not know how you get these categories, you need to say it}

On SWE-bench Pro (second block of Table~\ref{tab:rq2-dist-lflb-swe}), every representation still wins on real issues that the baseline fails ($20$--$28$ instances), and the win categories reflect each map's localization strength. For \textbf{AST}, the main win is a \emph{report-named in-file fault} ($18/28$): the issue names the faulty symbol, and the containment edge surfaces its implementation directly. For \textbf{callgraph}, it is a \emph{call-reachable cross-file fault} ($11/20$): the needed fix is in a caller or callee of the reported symbol, so the call edge exposes the coupled function that flat search can miss. For \textbf{deps}, it is a \emph{single-module dump} ($17/21$): one dependency-map call returns the target module and its test, giving the agent the relevant scaffold before editing. For \textbf{docs}, it is a \emph{cross-file shared-symbol bug} ($20/21$): a shared symbol, flag, or API contract links the affected files, and full-text search returns the related source together. For \textbf{hybrid}, the dominant win is \emph{lands the single target file} ($14/24$): the merged map resolves the report terms to the offending file, while the baseline gives up or lands elsewhere. The losses are more uniform. For every representation, the dominant failure is \emph{correct localization, wrong patch} ($16$--$23$ losses): the map finds the gold-patch files and the agent edits them, but the produced fix is logically wrong. Thus, unlike LinuxFLBench, SWE-bench Pro failures are often not navigation failures. Resolution remains close to the baseline ($44.3$--$50.6\%$ vs.\ $49.0\%$ in Table~\ref{tab:swebench}), suggesting that many wins and losses reflect one-shot patch variance rather than a stable representation effect. The remaining representation-independent failures are task shapes that no source-symbol map captures well: \emph{tree-wide find-and-replace}, where the index gives a definition but not all reference sites; \emph{new-file creation}, where there is no existing file to retrieve; and \emph{non-Python fix sites}, such as YAML or \texttt{docker-compose}. These cases defeat all maps and are therefore limitations of the task setting rather than of one representation. Figure~\ref{fig:eg-deps} gives a verified win/loss pair from the deps representation, both contrasted with the baseline.

\vspace{-0.05in}
\rescueregress
{fig:eg-deps}
{A matched deps win/loss pair on SWE-bench Pro%: one import-map query replaces the read phase when the fix lives in a named module (top), but the map cannot adjudicate the patch logic once the file is found, so a wrong $k{=}1$ attempt loses (bottom).}
}
{Deps win: single-module dump on a named target (SWE-bench Pro, ansible-3889ddeb)}
{\emph{Task:} add an argument to a single Ansible module. One \texttt{deps\_node} call on the named file dumps the whole target module and its test up front, handing the agent the complete argument-spec scaffold, so it edits the lone implementation file directly. The baseline reconstructs the same context through a long grep/Read march (roughly $80$ turns versus deps' $18$). \textbf{When the fix lives in one named module, one import-map query replaces the entire read phase.}}
{Deps loss: correct localization, wrong patch (SWE-bench Pro, ansible-942424e1)}
{\emph{Task:} fix an algorithm-/environment-dependent bug. \texttt{deps\_explore} reaches the exact fault file in a couple of calls (localization is never in doubt), but the single $k{=}1$ patch attempt is wrong, and the run fails on fix logic where the import map cannot adjudicate. The baseline, reaching the same file, happens to land a passing patch on its one attempt. \textbf{The map found the right file; the loss is in the edit, and at one run per instance it is fix-variance, not a localization failure.}}

%\song{For the performance, you did not show the other two benchmarks, please add them here}

\paragraph{Performance on security patching}
SecBench should not be treated as an editing-bound task because its input signal is stack-centered: prompts often expose crash stack frames, and success depends on following them to the edit location and execution context. This explains why \textbf{Callgraph} performs best. Its main win category is \emph{flow localization} ($9/18$): the callgraph query or selected node overlaps a crash-stack function or final edited file, helping the agent move from the crashing frame to caller/callee context that owns the vulnerable invariant, most often in parser, demuxer, decompression, and runtime paths. The comparison with \textbf{Hybrid}, which also contains call edges, shows that the advantage is not just more information but a more salient execution relation: among the 18 instances that Callgraph resolved but Hybrid did not, Hybrid edited the same file as Callgraph in 14 cases but failed to produce a successful patch. This suggests that reaching the relevant file was insufficient and the additional context in Hybrid may have made the execution-flow information less prominent. AST is often too local to the crash function, while Deps and Docs help only when names or comments expose the vulnerable path. SecBench favors Callgraph because many fixes require following caller/callee context from stack frames, but this does not make Callgraph universally best. Figure~\ref{fig:eg-callgraph-sec} gives a verified win/loss pair from the callgraph representation.

\vspace{-0.05in}
\rescueregress
{fig:eg-callgraph-sec}
{A matched callgraph win/loss pair on SecBench%: the call path turns the crash stack into the true repair site when the fix is call-reachable (top), but the call-centered view over-weights the crashing helper when the real fix is a caller-side format invariant (bottom).}
}
{Callgraph win: stack path to repair site (SecBench~\texttt{gpac.cve-2023-4682})}
{\emph{Vulnerability:} an AVI demuxing overflow whose sanitizer stack names \texttt{avi\_read}/\texttt{AVI\_read\_frame}. Callgraph's first query used \texttt{AVI\_read\_frame avi\_read avidmx\_process}, linking the crash helper to the demuxer path. The baseline added a size cap in \texttt{src/media\_tools/avilib.c}, but the sanitizer still fired. Callgraph instead produced a one-line state fix in \texttt{src/filters/dmx\_avi.c}, setting the AVI video position before reading the frame, and the patch resolved. \textbf{Here the call path turned the crash stack into the repair site rather than only the crashing helper.}}
{Callgraph loss: stack sink is not the invariant owner (SecBench~\texttt{faad2.cve-2021-32278})}
{\emph{Vulnerability:} a FAAD2 heap overflow with a stack rooted at \texttt{lt\_prediction} in \texttt{libfaad/lt\_predict.c}. Callgraph queried that sink and edited the same file, clamping a computed index; the patch compiled, but the sanitizer still fired. The baseline instead patched \texttt{libfaad/syntax.c}, propagating the return value from \texttt{ltp\_data} so invalid long-term-prediction metadata is rejected before prediction. \textbf{This is the failure mode: a call-centered view can make the crash helper too salient when the real fix is a caller-side format invariant.}}

\paragraph{Performance on test generation}
SWT-bench is different. Its dominant win and loss category is the same for every representation: the run writes tests in the same or neighboring test target as the baseline, and the $k{=}1$ generated test happens to reproduce, or not reproduce, the bug. Thus SWT-bench has small, mixed accuracy nets ($-11$ to $+5$) because the decisive step is authoring a discriminating assertion, not finding the repository neighborhood. The general lesson is therefore narrower than ``editing-bound tasks are noise'': representations help only when their relation matches the task's current bottleneck, and once the bottleneck becomes the patch invariant or the test assertion, map context can change trajectories but often becomes overhead.

\subsubsection{Root Causes for Token Benefit}

Applying the same protocol to the token axis, the \textbf{fewer-token} runs are dominated by one mechanism. On these instances, representations mostly help through \emph{read displacement}: the map returns relevant code bodies inline, so the agent treats them as already read and avoids many source-file reads. On \textbf{LinuxFLBench}, read displacement accounts for the majority of each representation's fewer-token runs (AST $79/111$, callgraph $109/109$, deps $72/115$, docs $93/115$, hybrid $85/117$). On \textbf{SWE-bench Pro}, it is again the largest single reason (read displacement explains 115–188 of each representation's 181–197 fewer-token runs). On \textbf{SecBench} and \textbf{SWT-bench}, the balance tips the other way: for every representation except the call graph on SWT-bench, more instances use more tokens than fewer (Table~\ref{tab:rq2-landscape}), because after the crash frame or test target is found, the agent must still read and edit, so map context is added on top of native browsing.

When representations use \textbf{more tokens} than the baseline, the overspend comes from a few recurring modes. \emph{Map overhead} is the main cost in the easy fault-localization cases: the baseline already solves the task cheaply, and the representation adds extra map calls for the same answer (e.g.\ AST $30/41$ of the LinuxFLBench more-token runs). The remaining more-token runs are \emph{mixed}, combining redundant re-exploration after the relevant file is found (most common for docs), large verbatim-source payloads from a single merged-map query (common for hybrid), and noisy imports pulled in by generic symbol names (common for deps). Thus, costs arise either from unavoidable map overhead on easy cases or from too many, too large, or too noisy map calls. Figure~\ref{fig:eg-docs} shows a verified Docs fewer-token/more-token pair.

\vspace{-0.05in}
\rescueregress
{fig:eg-docs}
{Docs token contrast on LinuxFLBench%: the text map displaces the read phase when comments anchor the fault (top), but its cheap queries become additive overhead when the file is already trivially named (bottom).}
}
{Docs fewer-token run: read displacement (LinuxFLBench~16043)}
{\emph{Bug:} a comment-/term-rich fault localization where docs answers from the full-text map alone. \texttt{docs\_explore} returns the fault file's source verbatim (``treat as already read''), so the agent ranks it correctly with essentially zero source reads, while the baseline reads $70$--$114$\,KB to reach the same answer. \textbf{The text map displaces the read phase entirely, for $0.15\times$ the baseline's tokens at equal accuracy.}}
{Docs more-token run: over-exploration of an already-found file (LinuxFLBench~11391)}
{\emph{Bug:} a report that names the file outright, which the baseline localizes in a handful of cheap actions. Docs' inexpensive verbatim-source query instead invites redundant refinement and big-result drilling on the \emph{already-found} file: the answer is right, but the exploration runs additively on top of what a plain search would have done, inflating tokens to $27.7\times$ the baseline. \textbf{When the file is trivially named, the map's cheap-explore loop becomes pure overhead.}}

\begin{findingbox}
\textbf{RQ2 takeaway.}
%\song{update the take away message, give an overview first, and then details}
Overall, representations help when their relation matches the task bottleneck. They separate most under \emph{localization} bottlenecks: AST helps named in-file faults, callgraph call-reachable fixes, deps import/API clusters, docs term-rich faults, and hybrid combined cues; the baseline remains stronger for out-of-map work such as occurrence enumeration, new-file creation, and config edits. Once the bottleneck moves past finding code, in repair, security patching, and test generation, only the relation exposing the needed edit or test still helps, and many remaining differences are one-shot patch/test variance. Token cost follows the same pattern: maps save tokens when they displace localization reads, and add cost when the agent must still read and edit.
\end{findingbox}

\section{Threats to Validity}
\label{sec:threats}

Several threats qualify our findings. \emph{Internally}, agent runs are not fully deterministic, so the same model and prompt may take different paths or patches. We mitigate this by fixing the agent stack, model, effort setting, prompt, and task IDs across workflows and by separating infrastructure failures from representation-content effects. However, representation workflows also add usage steering and, on repair, a pre-tool-use hook that forces one map query before native reads; the measured treatment therefore combines representation content, availability, and a map-first workflow, so behavior changes such as reduced native reads are not effects of the data structure alone. \emph{Externally}, we use one main agent stack and four repository-level benchmarks, so results may differ for other models, agents, languages, or repositories, and our claims concern these task families rather than all software engineering. \emph{For construct validity}, our metrics approximate the concepts of interest: benchmark success does not fully capture patch quality or developer usefulness, token and tool counts omit human and one-time infrastructure costs (map construction, local compute, clone/Docker setup, grading), and qualitative attribution requires judgment, so we assign representation effects only with direct trajectory evidence. \emph{Statistically}, we sample at a 95\% confidence level with a 5\% margin of error using the same task IDs across workflows, but each instance is run once per workflow ($k{=}1$), not under a multi-sample protocol such as pass@$k$~\cite{chen2021evaluating}. Small success differences, especially on SWT-bench and SecBench, may reflect one-shot patch- or test-generation variance; we therefore emphasize large, consistent effects and trajectory evidence over small rank changes and treat close resolution rates as descriptive.

\section{Conclusion}
\label{sec:conclusion}
We study how code representations affect agentic software engineering tasks. We expose AST, call graph, dependency, documentation, and hybrid repository views as MCP tools, and compare them with unstructured repository access across fault localization, program repair, test generation, and security patching.

Our results show that the benefits are task-dependent. Code representations help most when localization is the bottleneck: they improve fault localization, reduce token usage, and lower browsing effort in program repair without materially changing resolution rates. When tasks require more than localization, the trade-offs vary. Callgraph improves security patching and reduces token usage, but all representation workflows increase time cost. For test generation, gains are modest, while token and time costs increase.

Overall, software engineering agents should select code representations based on task bottlenecks rather than use a fixed representation strategy. Raw search remains useful for exact strings and non-code artifacts, while structural views are most useful when code relations guide localization.

\balance
\bibliographystyle{IEEEtran}
\bibliography{ref}

\end{document}